\documentclass[a4paper,12pt]{article}
\usepackage{graphicx}
\usepackage{amssymb}
\usepackage{amsmath}
\usepackage{cite}
\usepackage{axodraw4j}

\def\be{\begin{equation}}
\def\ee{\end{equation}}
\def\bea{\begin{eqnarray}}
\def\eea{\end{eqnarray}}
\def\beb{\begin{eqnarray*}}
\def\eeb{\end{eqnarray*}}
\def\pat{\partial}

\newlength{\myVSpace}
\begin{document}
\begin{center}
{\Large {\bf Higgs Couplings
            in NonCommutative Standard Model   }}
\end{center}
\vskip 1cm \centerline{ S. Batebi\footnote{s.batebi@ph.iut.ac.ir},
M. Haghighat\footnote{mansour@cc.iut.ac.ir}, S.
Tizchang\footnote{s.tizchang@ph.iut.ac.ir} and H.
Akafzadeh\footnote{h.akafzade@ph.iut.ac.ir} }\vskip .5cm
\centerline{\it
   Department of Physics, Isfahan University of Technology, Isfahan 84156-83111, Iran\\
 }
\begin{center}

{\bf Abstract}
\\

\medskip
\begin{minipage}{135mm}
{ We consider the Higgs and Yukawa parts of the Non-Commutative
Standard Model (NCSM).  We explore the NC-action to give all Feynman
rules for couplings of the Higgs boson to electro-weak gauge fields
and fermions.}
\end{minipage}
\end{center}
\medskip
\section{Introduction}
After discovery of the Higgs particle on 4 July 2012 at LHC
\cite{LHC}, it is tentatively confirmed to have zero spin and
positive parity \cite{spin}.  Nonetheless, it is yet to be
 determined if the particle discovered is the prediction of the
Standard Model, or whether, as predicted by the other theories
beyond the SM.  However, the observed properties of the Higgs boson,
such as its mass, couplings and decay rates, can constrain models
beyond SM.  In string theory, which is a candidate for explaining
gravity, the endpoints of an open string on D-branes in a constant
B-field background, live on a noncommutative space\cite{AASJ,SWM}.
Therefore, noncommutative field theory (NCFT) appears as a
low-energy limit of open string theory with constant antisymmetric
background field. The NC field theories and their phenomenological
aspects have been explored for many years \cite{NCFT}.  There are
two approaches to constructing the gauge theories in NC-space.  In
the first one, the gauge group is restricted to $U(n)$ and the
symmetry groups such as a $SU(n)$ can be achieved by an appropriate
symmetry breaking \cite{Un}.  In the second one, one can construct
directly a gauge theory based on a $SU(n)$ gauge group-and
consequently, the standard model-in noncommutative space
\cite{NCSM}.  After introducing the Feynman rules for the
noncommutative standard model (NCSM) in \cite{feyn} the
phenomenological aspects of NCSM have been considered by many
authors \cite{phenoNCSM}.  Here we would like to complete the
Feynman rules for the the Higgs and Yukawa sectors of the NCSM.  In
fact, in the both sectors there are many new couplings for the Higgs
boson with the other particles of the SM.  These new interactions
lead to many new interesting decay and production modes for the
Higgs particle.  Therefore, examining this part of NCSM can provide
new bounds on the noncommutative space-time.

In section 2 we give a brief review on the NCSM.  In section
\ref{sec:higgs} and \ref{Yukawa}, Feynman rule for the Higgs and
Yukawa part is derived, respectively. In Section 5, we summarize our
results and give concluding remarks.

\section{NCSM}
 In noncommutative space-time coordinates are operators which in the canonical version
 satisfy the following commutation relation
\begin{equation}
\left[\hat{x}^\mu,\hat{x}^\nu\right]=i\theta^{\mu\nu}=i\frac{C^{\mu\nu}}{\Lambda_{\tiny{NC}}^2},
\end{equation}
where $C^{\mu \nu}$ is a real antisymmetric tensor and
$\Lambda_{\tiny{NC}}$ is noncommutative scale. According to the
Weyl-Moyal correspondence, in a noncommutative field theory the
ordinary product between the fields should be replaced by the star
product which can be defined as\cite{star}
\begin{equation}
(f\star g)(x)=f(x)\exp(\frac{i}{2}\overleftarrow{\partial_\mu}
\theta^{\mu\nu}\overrightarrow {\partial_\nu})g(x). \label{a3}
\end{equation}
Using this correspondence, however, there are some problems to
construct non-Abelian gauge theories on noncommutative space. In
noncommutative QFT the field strength tensor is defined as follows
  \begin{equation}
  F_{\mu \nu}=[D_{\mu},D_{\nu}]_{\star}=\partial_{\mu} A_{\nu}-\partial_{\nu} A_{\mu}-i[A_{\mu},A_{\nu}]_{\star},
  \end{equation}
where the last term doesn't vanish even in the Abelian U(1) gauge
group.  In this framework the allowed particles have $\{0,\pm 1\}$
charges and quarks with the fractional charges can not be
accommodated in the model \cite{R4}. The second problem appears in
the commutator of two infinitesimal gauge transformations,
$\Lambda=\Lambda_a T^a$ and $\Lambda'=\Lambda'_b T^b$, where
\cite{Wess,gauge}
\begin{eqnarray}
[\Lambda,\Lambda']=\frac{1}{2}[T^a,T^b]\{\Lambda^a(x),\Lambda'^b(x)\}_{\star}+
\frac{1}{2}\{T^a,T^b\}[\Lambda^a(x),\Lambda'^b(x)]_{\star}.
\label{psir}
\end{eqnarray}
Non-vanishing anticommutator in the last term reproduce all the
higher powers of the generators in a non-Abelian gauge group such as
a $SU(N)$. It seems an enveloping algebra, consisting of all ordered
tensor powers of the generators, can be a proper choice to solve
this problem. Meanwhile, the infinite number of degrees of freedom
of the NC-gauge parameters and fields can be restricted demanding
that they depend on the algebra-valued quantities and their
space-time derivatives only. This requirement, based on the
equivalence between ordinary and noncommutative gauge fields to any
finite order in $\theta_{\mu\nu}$ and the existence of the
Seiberg-Witten map to all orders, leads to expansions for matter
field $\hat{\psi}$, gauge parameters $\hat{\Lambda}$, gauge fields
$\hat{A_\mu}$ and scalar field $\hat{\phi}$  as follows
\cite{SWM,gauge}

\begin{eqnarray}
\widehat{\psi}&=&\psi+\frac{1}{2}\theta^{\mu\nu}A_{\nu}\partial_{\mu}\psi +\frac{i}{8}\theta^{\mu\nu}[A_{\mu},A_{\nu}] +\mathcal{O}(\theta^2),\\
\nonumber
\widehat{A}_{\mu}&=&A_{\mu}+\frac{1}{4}\theta^{\rho\nu}\{A_{\nu},(\partial_{\rho}A_{\mu}-F_{\rho\mu})\}+\mathcal{O}(\theta^2),\\
\nonumber
\widehat{\Lambda}&=&\Lambda+\frac{1}{4}\theta^{\mu\nu}\{A_{\nu},\partial_{\mu}\Lambda\}+\mathcal{O}(\theta^2),\\
\nonumber
  \widehat{\phi}&=& \phi +
  \frac{1}{2} \theta^{\mu\nu}A_{\nu}
      \left( \partial_{\mu} \phi
       - \frac{i}{2} (A_{\mu} \phi -\phi A'_{\mu})\right)
   \\ \nonumber&&
  +\frac{1}{2} \, \theta^{\mu\nu}
  \left( \partial_{\mu} \phi
       - \frac{i}{2} (A_{\mu}\phi -\phi A'_{\mu}) \right)
       A'_{\nu}+\mathcal{O}(\theta^2),
\label{gaie}
\end{eqnarray}
where the scalar field can be transformed on the left and right
under two different gauge groups with the corresponding gauge fields
$A$ and $A'$.

Now the NC-standard model based on the gauge group
  $SU(3)_C \times SU(2)_L\times U(1)$, can be constructed in two steps:
\begin{enumerate}
 \item  Replacing the ordinary products and fields in the action of the standard model with the star
products and the corresponding NC-fields, respectively.
 \item Substituting the noncommutative fields for each corresponding
commutative one via the Seiberg-Witten map given in (5).
 \end{enumerate}
 The minimal NCSM based on the Seiberg-Witten maps has been
 introduced in \cite{NCSM}.  The action of noncommutative
standard model can be separated into four parts as
\begin{eqnarray}
S_{\mbox{\tiny NCSM}}= S_{\mbox{\tiny fermions}} + S_{\mbox{\tiny
gauge}} + S_{\mbox{\tiny{Higgs}}}+ S_{\mbox{\tiny{Yukawa}}} \, .
\label{eq:Sncsm}
\end{eqnarray}
Meanwhile, the Feynman rules for $S_{\mbox{\tiny fermions}}$ and
$S_{\mbox{\tiny gauge}}$ is fully introduced in \cite{feyn}.  In the
next sections we explore the remaining parts to find the Feynman
rules for the Higgs interactions.                                                                                                                           
\section{\label{sec:higgs}Higgs part of the NCSM action}
The Higgs part of the NCSM action up to the first order of $\theta$
can be written as follows\cite{feyn}:
\begin{eqnarray}
S_{\mbox{\tiny{Higgs}}} {\hspace{-1mm}}&=&
 \int d^4x \bigg( ({\bf D}_\mu \Phi)^\dagger
 ({\bf D}^\mu \Phi)
- \mu^2 {\Phi}^\dagger  \Phi
- \lambda \,
(\Phi^\dagger  \Phi)^2
\bigg)
\nonumber \\ &&
+ \frac{1}{2} \, \theta^{\alpha\beta}\int {\hspace{-1mm}}d^4 x\,
\Phi^\dagger \Bigg( {\bf U}_{\alpha
    \beta} + {\bf U}^\dagger_{\alpha \beta}
    + \frac12  \mu^2 \, {\bf F}_{\alpha\beta} -
    2 i\lambda\, \Phi ({\bf D}_\alpha\Phi)^\dagger
    {\bf D}_\beta \Bigg)\Phi \,,
\nonumber \\ &&
\label{eq:H}
\end{eqnarray}
where ${\bf D}_\mu=  \partial_{\mu} {\bf 1}-i {\bf V}_{\mu}$ and
${\bf F}_{\mu \nu}= \partial_{\mu} {\bf V}_{\nu}-\partial_{\nu} {\bf
V}_{\mu} -i [{\bf V}_{\mu},{\bf V}_{\nu}]$ are the ordinary
covariant derivative and field strength tensor, respectively, and
\begin{eqnarray}
{\bf U} _{\alpha \beta} &=& \left(
    \stackrel{\leftarrow}{{\bf \pat}^\mu}
    + \, i{\bf V}^\mu \right) \Bigg(
    -\, {\bf \pat}_\mu {\bf V}_\alpha \, {\bf \pat}_\beta
    - \, {\bf V}_\alpha {\bf \pat}_\mu{\bf \pat}_\beta
    + \, {\bf \pat}_\alpha {\bf V}_\mu {\bf \pat}_\beta
 \nonumber \\ & &
    + \, i {\bf V}_\mu {\bf V}_\alpha {\bf \pat}_\beta
    + \, \frac{i}2 {\bf V}_{\alpha}{\bf V}_{\beta}{\bf \pat}_{\mu}
+ \, \frac{i}2 {\bf \pat}_{\mu}({\bf V}_{\alpha}{\bf V}_{\beta})
  \nonumber \\ & &
+ \, \frac12 {\bf V}_{\mu}{\bf V}_{\alpha}{\bf V}_{\beta} + \,
\frac{i}2 \{{\bf V}_\alpha,\, {\bf \pat}_\beta {\bf V}_\mu + {\bf
F}_{\beta\mu} \} \Bigg), \label{aa}
\end{eqnarray}
in which ${\bf 1}$ is a unit matrix and the gauge field ${\bf
V}_{\mu}=g' {\mathcal A}_\mu Y_{\Phi} {\bf 1} + g B_\mu^a T_L^a$ can
be easily rewritten as
\begin{equation}
{\bf V}_\mu = \left( \begin{array}{cc}
\displaystyle
g'\mathcal A_\mu Y_{\Phi} + g T_{3,\phi_{\mbox{\tiny up}}}
    B_\mu^3
    &
\displaystyle
\frac{g}{\sqrt{2}} W_{\mu}^+\\[0.4cm]
\displaystyle
    \frac{g}{\sqrt{2}} W_{\mu}^- &
\displaystyle
g'\mathcal A_\mu Y_{\Phi} + g T_{3,\phi_{\mbox{\tiny down}}}
    B_\mu^3 \end{array} \right)
\, ,
\label{eq:V}
\end{equation}
where $Y_{\Phi}=1/2$, $T_{3,\phi_{\mbox{\tiny up}}}=1/2$,
$T_{3,\phi_{\mbox{\tiny down}}}=-1/2$.  Consequently, in terms of
the physical fields $Z_{\mu}$ and $A_{\mu}$,
\begin{eqnarray}
Z_{\mu} &=&
\frac{-g' {\cal A}_{\mu} + g B_{\mu}^3}{\sqrt{g^2+{g'}^2}}
        = -\sin \theta_W {\cal A}_{\mu}
            + \cos \theta_W B_{\mu}^3
        \, ,
\nonumber\\
A_{\mu} &=& \frac{g {\cal A}_{\mu} + g' B_{\mu}^3}{\sqrt{g^2+{g'}^2}}
        = \cos \theta_W {\cal A}_{\mu}
            + \sin \theta_W B_{\mu}^3
\, ,
\label{eq:Amu}
\end{eqnarray}
the diagonal components of the gauge field ${\bf V}_\mu$ can be
obtained as follows
\begin{eqnarray}
V_{11,\mu} &=& e A_{\mu}
   + \frac{g}{2 \cos \theta_W} (1-2 \sin^2 \theta_W) Z_{\mu}
\, ,
\nonumber \\
V_{22,\mu} &=& - \frac{g}{2 \cos \theta_W} Z_{\mu}
\, .
\label{eq:V1122}
\end{eqnarray}
Meanwhile, the Higgs field in the unitary gauge is $ \phi(x) =
\frac1{\sqrt{2}} \left(\begin{array}{c}0\\ h(x) +
v\end{array}\right)$, where $v=\sqrt{-\mu^2/\lambda}$ is the Higgs
vacuum expectation value and $h(x)$ is the physical Higgs field. Now
the Higgs action (\ref{eq:H}), by using (\ref{aa}) to
(\ref{eq:V1122}), can be rewritten in terms of the physical fields
to obtain all Lagrangian densities containing the Higgs interactions
with the gauge fields as follows
\begin{itemize}
\item

\begin{itemize}
\item one Higgs 2 neutral gauge bosons, $hZZ$:

\begin{eqnarray}
{\cal L}_{hZZ}&=& {\cal L}^{SM}_{hZZ}+ {\cal
O}(\theta^2).\label{hzz}
\end{eqnarray}

\item 2 Higgs's 2 neutral gauge bosons, $hhZZ$:

\begin{eqnarray}
{\cal L}_{hhZZ}&=& {\cal L}^{SM}_{hhZZ}+ {\cal
O}(\theta^2).\label{hhzz}
\end{eqnarray}

\item 2 Higgs's a neutral gauge boson, $hhZ$:

\begin{eqnarray}
{\cal L}_{hhZ}&=&(\frac{g}{4\cos\theta_{w}})\theta^{\alpha\beta}
[-(\partial_{\mu}\partial^{\mu}h)(\partial_{\beta}h)Z_{\alpha}
\nonumber\\&&+
(\partial_{\alpha}\partial^{\mu}h)(\partial_{\beta}h)Z_{\mu}+2\lambda
v^{2}h(\partial_{\alpha}h)Z_{\beta}].\label{hhz}
\end{eqnarray}

\item 3 Higgs's a neutral gauge boson, $hhhZ$:

\begin{eqnarray}
{\cal L}_{hhhZ}=(\frac{3g}{4\cos\theta_{w}})\theta^{\alpha\beta}
\lambda v h^{2}(\partial_{\alpha}h)Z_{\beta}.\label{hhhz}
\end{eqnarray}

\item 4 Higgs's a neutral gauge boson, $hhhhZ$:

\begin{eqnarray}
{\cal L}_{hhhhZ}=(\frac{g}{4\cos\theta_{w}})\theta^{\alpha\beta}
\lambda h^{3}(\partial_{\alpha}h)Z_{\beta}.\label{hhhhz}
\end{eqnarray}

\item one Higgs 2 charged gauge bosons, $hW^{+}W^{-}$:

\begin{eqnarray}
{\cal
L}_{hW^{+}W^{-}}&=&\frac{g^{2}}{2}vhW^{-\mu}W_{+\mu}\nonumber\\&&+\frac{ig^{2}}{8}v\theta^{\alpha\beta}
[-(\partial_{\beta}h)(W^{-\mu}(\partial_{\mu}W^{+}_{\alpha})-W^{+\mu}(\partial_{\mu}W^{-}_{\alpha})
\nonumber\\&& -
W^{-\mu}(\partial_{\alpha}W^{+}_{\mu})+W^{+\mu}(\partial_{\alpha}W^{-}_{\mu}))-(\partial_{\mu}\partial_{\beta}h)(W^{-\mu}W^{+}_{\alpha}
\nonumber\\&&-W^{+\mu}W^{-}_{\alpha})+
(\partial^{\mu}h)((\partial_{\mu}W^{-}_{\alpha})W^{+}_{\beta}-(\partial_{\mu}W^{+}_{\alpha})W^{-}_{\beta})
\nonumber\\&& +2\lambda v^{2} h
W^{+}_{\alpha}W^{-}_{\beta}].\label{hww}
\end{eqnarray}

\item 2 Higgs's 2 charged gauge bosons, $hhW^{+}W^{-}$:
\begin{eqnarray}
{\cal
L}_{hhW^{+}W^{-}}&=&\frac{g^{2}}{4}h^{2}W^{+\mu}W^{-}_{\mu}\nonumber\\&&+\frac{ig^{2}}{8}\theta^{\alpha\beta}
[-h(\partial_{\beta}h)(W^{-\mu}(\partial_{\mu}W^{+}_{\alpha})-W^{+\mu}(\partial_{\mu}W^{-}_{\alpha})\nonumber\\&&-
W^{-\mu}(\partial_{\alpha}W^{+}_{\mu})
+W^{+\mu}(\partial_{\alpha}W^{-}_{\mu}))\nonumber\\&&
-h(\partial_{\mu}\partial_{\beta}h)(W^{-\mu}W^{+}_{\alpha}-W^{+\mu}W^{-}_{\alpha})\nonumber\\&&+
h(\partial^{\mu}h)
((\partial_{\mu}W^{-}_{\alpha})W^{+}_{\beta}-(\partial_{\mu}W^{+}_{\alpha})W^{-}_{\beta})\nonumber\\&&
+(\partial^{\mu}h)(\partial_{\beta}h)(W^{-}_{\mu}W^{+}_{\alpha}-W^{+}_{\mu}W^{-}_{\alpha})
\nonumber\\&& +
(\partial^{\mu}h)(\partial_{\mu}h)W^{-}_{\alpha}W^{+}_{\beta}+5\lambda
v^{2} h^{2}W^{+}_{\alpha}W^{-}_{\beta}].\label{hhww}
\end{eqnarray}

\item 3 Higgs's 2 charged gauge bosons, $hhhW^{+}W^{-}$:

\begin{eqnarray}
{\cal L}_{hhhW^{+}W^{-}}=\frac{i
g^{2}}{2}v\lambda\theta^{\alpha\beta}
h^{3}W^{+}_{\alpha}W^{-}_{\beta}.\label{hhhww}
\end{eqnarray}

\item 4 Higgs's 2 charged gauge bosons, $hhhhW^{+}W^{-}$:

\begin{eqnarray}
{\cal
L}_{hhhhW^{+}W^{-}}=\frac{ig^{2}}{8}\lambda\theta^{\alpha\beta}
h^{4}W^{+}_{\alpha}W^{-}_{\beta}.\label{hhhhww}
\end{eqnarray}

\item one Higgs 2 charged gauge bosons one photon, $hW^{+}W^{-}\gamma$:

\begin{eqnarray}
{\cal L}_{hW^{+}W^{-}A}&=&\frac{eg^{2}}{8} v\theta^{\alpha\beta}
[(-(\partial_{\beta}h)A^{\mu}\nonumber\\&&+2h(\partial^{\mu}A_{\beta}-\partial_{\beta}A^{\mu}))
(W^{-}_{\mu}W^{+}_{\alpha}+W^{+}_{\mu}W^{-}_{\alpha})\nonumber\\&&+2h(\partial_{\beta}A_{\alpha})W^{+\mu}W^{-}_{\mu}].\label{hwwg}
\end{eqnarray}

\item 2 Higgs's 2 charged gauge bosons one photon, $hhW^{+}W^{-}\gamma$:

\begin{eqnarray}
{\cal L}_{hhW^{+}W^{-}A}&=&\frac{e g^{2}}{8}\theta^{\alpha\beta}
[(-h(\partial_{\beta}h)A^{\mu}\nonumber\\&&+h^{2}(\partial^{\mu}A_{\beta}-\partial_{\beta}A^{\mu}))
(W^{-}_{\mu}W^{+}_{\alpha}+W^{+}_{\mu}W^{-}_{\alpha})\nonumber\\&&+h^{2}(\partial_{\beta}A_{\alpha})W^{+\mu}W^{-}_{\mu}].\label{hhwwg}
\end{eqnarray}

\item one Higgs 3 neutral gauge bosons, $hZZZ$:

\begin{eqnarray}
{\cal L}_{hZZZ}&=&\frac{1}{16}(\frac{g}{\cos\theta_{w}})^{3}v
\theta^{\alpha\beta} [(\partial_{\beta}h)(Z^{\mu}Z_{\mu}Z_{\alpha})
\nonumber\\&&
+2hZ^{\mu}Z_{\alpha}(2\partial_{\beta}Z_{\mu}-\partial_{\mu}Z_{\beta})].\label{hzzz}
\end{eqnarray}

\item 2 Higgs's 3 neutral gauge bosons, $hhZZZ$:

\begin{eqnarray}
{\cal
L}_{hhZZZ}&=&\frac{1}{16}(\frac{g}{\cos\theta_{w}})^{3}\theta^{\alpha\beta}
[h(\partial_{\beta}h)(Z^{\mu}Z_{\mu}Z_{\alpha})
\nonumber\\&&+h^{2}Z^{\mu}Z_{\alpha}(2\partial_{\beta}Z_{\mu}-\partial_{\mu}Z_{\beta})].\label{hhzzz}
\end{eqnarray}

\item one Higgs 2 charged gauge bosons one neutral gauge boson, $hW^{+}W^{-}Z$:

\begin{eqnarray}
\hspace{1.5cm}
 \mathcal{L}_{hW^{+}W^{-}Z}&=& \frac{g^3v}{8cos\theta_{W}}\theta^{\alpha\beta}\{(sin^2{\theta_{W}}Z_{\mu}(W^{-\mu}W^{+}_{\alpha}+W^{+\mu}W^{-}_{\alpha})
  \nonumber\\ &&+Z_{\alpha}W^{-\mu}W^{+}_{\mu}
 )(\partial_{\beta}h)
\nonumber\\
&&-h[Z^{\mu}(W^{+}_{\mu}(\partial_{\beta}W^{-}
_{\alpha})+W^{-}_{\mu}(\partial_{\beta}W^{+}_{\alpha}))
\nonumber\\
&&+(Z^{\mu}W^{+}_{\alpha}+Z_{\alpha}W^{+\mu})
(\partial_{\mu}W^{-}_{\beta}-\partial_{\beta}W^{-}_{\mu})\nonumber\\
&&
+(Z^{\mu}W^{-}_{\alpha}+Z_{\alpha}W^{-\mu})(\partial_{\mu}W^{+}_{\beta}-\partial_{\beta}W^{+}_{\mu})
 \nonumber\\ &&
 -cos2\theta_{W}[(W^{+\mu}W^{-}_{\alpha}+W^{-\mu}W^{+}_{\alpha})(\partial_{\mu}Z_{\beta}-\partial_{\beta}Z_{\mu})
  \nonumber\\ &&
 +(\partial_{\beta}Z_{\alpha})W^{+\mu}W^{-}_{\mu}]]\}.
 \label{hwwz}
\end{eqnarray}

\item 2 Higgs's 2 charged gauge bosons one neutral gauge boson, $hhW^{+}W^{-}Z$:

\begin{eqnarray}
\hspace{1.5cm}
 \mathcal{L}_{hhW^{+}W^{-}Z}&=& \frac{g^3}{8cos\theta_{W}}\theta^{\alpha\beta}\{(sin^2{\theta_{W}}Z_{\mu}(W^{-\mu}W^{+}_{\alpha}+W^{+\mu}W^{-}_{\alpha})
\nonumber\\ && +Z_{\alpha}W^{-\mu}W^{+}_{\mu}
 )(h\partial_{\beta}h)
  \nonumber\\ &&
-h^{2}[Z^{\mu}(W^{+}_{\mu}(\partial_{\beta}W^{-}
_{\alpha})+W^{-}_{\mu}(\partial_{\beta}W^{+}_{\alpha})) \nonumber\\
&&+(Z^{\mu}W^{+}_{\alpha}+Z_{\alpha}W^{+\mu})
(\partial_{\mu}W^{-}_{\beta}-\partial_{\beta}W^{-}_{\mu})\nonumber\\
&&
+(Z^{\mu}W^{-}_{\alpha}+Z_{\alpha}W^{-\mu})(\partial_{\mu}W^{+}_{\beta}-\partial_{\beta}W^{+}_{\mu})
 \nonumber\\ &&
 -cos2\theta_{W}[(W^{+\mu}W^{-}_{\alpha}+W^{-\mu}W^{+}_{\alpha})(\partial_{\mu}Z_{\beta}-\partial_{\beta}Z_{\mu})
  \nonumber\\ &&
 +(\partial_{\beta}Z_{\alpha})W^{+\mu}W^{-}_{\mu}]]\}.
 \label{hhwwz}
\end{eqnarray}
\\
\\

\item one Higgs 2 charged gauge bosons one neutral gauge boson one photon, $hW^{+}W^{-}Z\gamma$:

\begin{eqnarray}
{\cal L}_{hW^{+}W^{-}ZA}&=&
(-\frac{ievg^{3}}{8\cos\theta_{w}})\theta^{\alpha\beta}h
[2Z^{\mu}A_{\alpha}W^{-}_{\mu}W^{+}_{\beta}-2Z^{\mu}A_{\alpha}W^{+}_{\mu}W^{-}_{\beta}
\nonumber\\&&
+2Z^{\mu}A_{\mu}W^{-}_{\beta}W^{+}_{\alpha}+Z_{\beta}A_{\mu}W^{-\mu}W^{+}_{\alpha}\nonumber\\
&&-Z_{\beta}A_{\mu}W^{+\mu}W^{-}_{\alpha}].\label{hwwzg}
\end{eqnarray}

\item 2 Higgs's 2 charged gauge bosons one neutral gauge boson one photon, $hhW^{+}W^{-}Z\gamma$:

\begin{eqnarray}
{\cal
L}_{hhW^{+}W^{-}ZA}&=&(-\frac{ieg^{3}}{16\cos\theta_{w}})\theta^{\alpha\beta}h^{2}
[2Z^{\mu}A_{\alpha}W^{-}_{\mu}W^{+}_{\beta}-2Z^{\mu}A_{\alpha}W^{+}_{\mu}W^{-}_{\beta}
\nonumber\\&&
+2Z^{\mu}A_{\mu}W^{-}_{\beta}W^{+}_{\alpha}+Z_{\beta}A_{\mu}W^{-\mu}W^{+}_{\alpha}\nonumber\\
&&-Z_{\beta}A_{\mu}W^{+\mu}W^{-}_{\alpha}].\label{hhwwzg}
\end{eqnarray}

\item one Higgs 2 charged gauge bosons 2 neutral gauge bosons, $hW^{+}W^{-}ZZ$:

\begin{eqnarray}
{\cal L}_{hW^{+}W^{-}ZZ}&=&-iv
(\frac{g^{2}}{4\cos\theta_{w}})^{2}\theta^{\alpha\beta}h\{(1-2\sin^{2}\theta_{W})
[-Z^{\mu}Z_{\beta}W^{-}_{\mu}W^{+}_{\alpha} \nonumber\\&&
+Z^{\mu}Z_{\beta}W^{-}_{\alpha}W^{+}_{\mu}+2Z^{\mu}Z_{\mu}W^{-}_{\beta}W^{+}_{\alpha}]
+3Z^{\mu}[Z_{\alpha}W^{-}_{\mu}W^{+}_{\beta} \nonumber\\&&
+Z_{\beta}W^{-}_{\alpha}W^{+}_{\mu}-Z_{\mu}W^{-}_{\alpha}W^{+}_{\beta}]\}.\label{hwwzz}
\end{eqnarray}

\item 2 Higgs's 2 charged gauge bosons 2 neutral gauge bosons, $hhW^{+}W^{-}ZZ$:

\begin{eqnarray}
{\cal L}_{hhW^{+}W^{-}ZZ}&=&\frac{-i}{2}
(\frac{g^{2}}{4\cos\theta_{w}})^{2}\theta^{\alpha\beta}h^{2}\{(1-2\sin^{2}\theta_{W})
[-Z^{\mu}Z_{\beta}W^{-}_{\mu}W^{+}_{\alpha} \nonumber\\&&
+Z^{\mu}Z_{\beta}W^{-}_{\alpha}W^{+}_{\mu}+2Z^{\mu}Z_{\mu}W^{-}_{\beta}W^{+}_{\alpha}]
+3Z^{\mu}[Z_{\alpha}W^{-}_{\mu}W^{+}_{\beta} \nonumber\\&&
+Z_{\beta}W^{-}_{\alpha}W^{+}_{\mu}-Z_{\mu}W^{-}_{\alpha}W^{+}_{\beta}]\}.\label{hhwwzz}
\end{eqnarray}

\item one Higgs 4 charged gauge bosons, $hW^{+}W^{+}W^{-}W^{-}$:

\begin{eqnarray}
{\cal
L}_{hW^{+}W^{+}W^{-}W^{-}}=\frac{ig^{4}v}{8}\theta^{\alpha\beta}h
[W^{-\mu}W^{+}_{\mu}W^{-}_{\alpha}W^{+}_{\beta}].\label{hwwww}
\end{eqnarray}

\item 2 Higgs's 4 charged gauge bosons, $hhW^{+}W^{+}W^{-}W^{-}$:

\begin{eqnarray}
{\cal
L}_{hhW^{+}W^{+}W^{-}W^{-}}=\frac{ig^{4}}{16}\theta^{\alpha\beta}
h^{2}
[W^{-\mu}W^{+}_{\mu}W^{-}_{\alpha}W^{+}_{\beta}].\label{hhwwww}
\end{eqnarray}
\end{itemize}
\subsection{\label{G}{Gauge Terms}}
In this subsection, we list Feynman rules for the NCSM Higgs-gauge
couplings up to the first order of $\theta$.  To derive the rules,
it is assumed that all particles momenta are incoming into the
vertices.\\
\\
Equation (\ref{hhz}):
\item \hspace{-1cm}
\\
\\
\hspace{-1cm}
\begin{picture}(55,45) (30,-30)
\SetWidth{0.5}
\DashLine(30,-30)(50,-10){3}
\DashLine(50,-10)(30,11){3}
\Photon(80,-10)(50,-10){2}{7}
\Vertex(50,-10){1.5}
\Text(30,11)[lb]{$H(k)$}
\Text(30,-45)[lb]{$H(k')$}
\Text(75,-5)[lb]{$Z_{\rho}$}
\end{picture}
\\
\\
\\
\\
\hspace{-1cm}
\begin{equation}
-\frac{M_{Z}}{2v}[(k'^{2}k_{\beta}+k^{2}k'_{\beta})\theta^{\beta\rho}+
k'_{\alpha}k_{\beta}(k'-k)^{\rho}\theta^{\alpha\beta}-m_{H}^{2}(k'+k)_{\alpha}\theta^{\alpha\rho}],
\label{hhz1}
\end{equation}
\\
\\
\\
Equation (\ref{hhhz}):
\\
\\
\item \hspace{-1cm}
\\
\\
\begin{picture}(55,45) (30,-30)
\SetWidth{0.5}
\DashLine(30,-30)(50,-10){3}
\DashLine(20,-10)(50,-10){3}
\DashLine(30,10)(50,-10){3}
\Photon(80,-10)(50,-10){2}{7}
\Vertex(50,-10){1.5}
\Text(30,11)[lb]{$H(k)$}
\Text(-15,-10)[lb]{$H(k')$}
\Text(30,-43)[lb]{$H(k'')$}
\Text(75,-5)[lb]{$Z_{\rho}$}
\end{picture}
\vspace{-1.0cm}
\begin{equation}
\frac{3}{2}\frac{M_{Z}m_{H}^{2}}{v^{2}}\theta^{\alpha\rho}[k+k'+k'']_{\alpha},
\end{equation}
\\
\\
\\
Equation (\ref{hhhhz}):
\item \hspace{-1cm}
\\
\\
\begin{picture}(55,45) (30,-30)
\SetWidth{0.5}
\DashLine(30,-30)(50,-10){3}
\DashLine(20,-20)(50,-10){3}
\DashLine(20,0)(50,-10){3}
\DashLine(30,10)(50,-10){3}
\Photon(80,-10)(50,-10){2}{7}
\Vertex(50,-10){1.5}
\Text(30,11)[lb]{$H(k)$}
\Text(-15,-3)[lb]{$H(k')$}
\Text(-15,-27)[lb]{$H(k'')$}
\Text(30,-43)[lb]{$H(k''')$}
\Text(75,-5)[lb]{$Z_{\rho}$}
\end{picture}
\vspace{-1.50cm}
\begin{equation}
\frac{3}{2}\frac{M_{Z}m_{H}^{2}}{v^{3}}\theta^{\alpha\rho}[k+k'+k''+k''']_{\alpha},
\end{equation}
\\
\\
\\
Equation (\ref{hww}):
\\
\\
\item \hspace{-1cm}
\\
\\
\begin{picture}(55,45) (30,-30)
\SetWidth{0.5}
\DashLine(17,-10)(50,-10){3}
\Photon(75,10)(50,-10){2}{7}
\Photon(75,-30)(50,-10){2}{7}
\Vertex(50,-10){1.5}
\Text(7,-5)[lb]{$H(k)$}
\Text(75,8)[lb]{$W^+_{\gamma}(k_1)$}
\Text(75,-44)[lb]{$W^-_{\sigma}(k_2)$}
\end{picture}
\\
\begin{eqnarray}
&&
\hspace{1.5cm}
\frac{M_{W}^{2}}{2v}\{4ig^{\sigma\gamma}+k.(k_{1}+k_{2})\theta^{\sigma\gamma}+k_{\beta}(k_{2}^{\gamma}\theta^{\sigma\beta}-
k_{1}^{\sigma}\theta^{\gamma\beta})
\nonumber
\\
&& \hspace{1.1cm}
+k_{\beta}(k^{\gamma}\theta^{\sigma\beta}-k^{\sigma}\theta^{\gamma\beta})+
k_{\beta}(k_{1}-k_{2})_{\alpha}g^{\gamma\sigma}\theta^{\alpha\beta}-m_{H}^{2}\theta^{\gamma\sigma}\},
\end{eqnarray}
\\
\\
Equation (\ref{hhww}):
\item \hspace{-1cm}
\\
\\
\begin{picture}(55,45) (30,-30)
\SetWidth{0.5}
\DashLine(30,10)(50,-10){3}
\DashLine(30,-30)(50,-10){3}
\Photon(70,10)(50,-10){2}{7}
\Photon(70,-30)(50,-10){2}{7}
\Vertex(50,-10){1.5}
\Text(20,11)[lb]{$H(k)$}
\Text(20,-45)[lb]{$H(k')$}
\Text(70,10)[lb]{$W^+_{\gamma}(k_1)$}
\Text(70,-46)[lb]{$W^-_{\sigma}(k_2)$}
\end{picture}
\\
\begin{eqnarray}
&&
\hspace{1.5cm}
\frac{M_{W}^{2}}{2v^2}\{4ig^{\sigma\gamma}+(k'-k)_{\beta}[(k-k')^{\sigma}\theta^{\gamma\beta}-(k-k')^{\gamma}\theta^{\sigma\beta}]
\nonumber
\\
&&
\hspace{1.5cm}
-k_{1}^{\sigma}(k+k')_{\beta}\theta^{\gamma\beta}+k_{2}^{\gamma}(k+k')_{\beta}\theta^{\sigma\beta}+(k+k')_{\beta}(k_{1}-k_{2})_{\alpha}\
\nonumber
\\
&& \hspace{1.5cm} \times
g^{\gamma\sigma}\theta^{\alpha\beta}+[2k.k'+(k+k').(k_{1}+k_{2})]\theta^{\sigma\gamma}-5m_{H}^{2}\theta^{\gamma\sigma}\},
\end{eqnarray}
\\
\\
Equation (\ref{hhhww}):
\item \hspace{-1cm}
\\
\\
\begin{picture}(55,45) (30,-30)
\SetWidth{0.5}
\DashLine(30,10)(50,-10){3}
\DashLine(20,-10)(50,-10){3}
\DashLine(30,-30)(50,-10){3}
\Photon(70,10)(50,-10){2}{7}
\Photon(70,-30)(50,-10){2}{7}
\Vertex(50,-10){1.5}
\Text(25,13)[lb]{$H$}
\Text(7,-15)[lb]{$H$}
\Text(25,-42)[lb]{$H$}
\Text(68,7)[lb]{$W^+_{\gamma}$}
\Text(68,-46)[lb]{$W^-_{\sigma}$}
\end{picture}
\vspace{-1cm} 
\begin{equation}
6\frac{m_{H}^{2}M_{W}^{2}}{v^{3}}\theta^{\sigma\gamma},
\end{equation}
\\
\\
\\
\\
Equation (\ref{hhhhww}):
\item \hspace{-1cm}
\\
\\
\begin{picture}(55,45) (30,-30)
\SetWidth{0.5}
\DashLine(30,10)(50,-10){3}
\DashLine(20,-3)(50,-10){3}
\DashLine(20,-17)(50,-10){3}
\DashLine(30,-30)(50,-10){3}
\Photon(70,10)(50,-10){2}{7}
\Photon(70,-30)(50,-10){2}{7}
\Vertex(50,-10){1.5}
\Text(23,14)[lb]{$H$}
\Text(7,-5)[lb]{$H$}
\Text(7,-22)[lb]{$H$}
\Text(23,-43)[lb]{$H$}
\Text(70,8)[lb]{$W^+_{\gamma}$}
\Text(70,-44)[lb]{$W^-_{\sigma}$}
\end{picture}
\\
\vspace{-1cm} 
\begin{equation}
6\frac{m_{H}^{2}M_{W}^{2}}{v^{4}}\theta^{\sigma\gamma},
\end{equation}
\\
\\
Equation (\ref{hwwg}):
\item \hspace{-1cm}
\\
\\
\begin{picture}(55,45) (30,-30)
\SetWidth{0.5}
\DashLine(25,-10)(50,-10){3}
\Photon(70,10)(50,-10){2}{7}
\Photon(80,-10)(50,-10){2}{4}
\Photon(70,-30)(50,-10){2}{7}
\Vertex(50,-10){1.5}
\Text(8,-5)[lb]{$H(k)$}
\Text(75,8)[lb]{$W^+_{\gamma}(k_1)$}
\Text(83,-15)[lb]{$A_{\rho}(k_3)$}
\Text(75,-44)[lb]{$W^-_{\sigma}(k_2)$}
\end{picture}
\vspace{1cm}
\begin{eqnarray}
&&
\hspace{1.5cm}
e\frac{M_{W}^{2}}{2v}\{k_{\beta}(\theta^{\beta\gamma}g^{\sigma\rho}+\theta^{\beta\sigma}g^{\gamma\rho})+2(k_{3}^{\gamma}\theta^{\sigma\rho}\,
\nonumber
\\
&& \hspace{1.5cm}
+k_{3}^{\sigma}\theta^{\gamma\rho})+2k_{3\beta}(\theta^{\rho\beta}g^{\sigma\gamma}+\theta^{\beta\gamma}g^{\sigma\rho}+\theta^{\beta\sigma}g^{\gamma\rho})\},
\end{eqnarray}
\\
\\
Equation (\ref{hhwwg}):
\item \hspace{-1cm}
\\
\\
\begin{picture}(55,45) (30,-30)
\SetWidth{0.5}
\DashLine(30,10)(50,-10){3}
\DashLine(30,-30)(50,-10){3}
\Photon(70,10)(50,-10){2}{7}
\Photon(80,-10)(50,-10){2}{4}
\Photon(70,-30)(50,-10){2}{7}
\Vertex(50,-10){1.5}
\Text(15,13)[lb]{$H(k)$}
\Text(15,-47)[lb]{$H(k')$}
\Text(75,8)[lb]{$W^+_{\gamma}(k_1)$}
\Text(83,-15)[lb]{$A_{\rho}(k_3)$}
\Text(75,-44)[lb]{$W^-_{\sigma}(k_2)$}
\end{picture}
\vspace{.5cm}
\begin{eqnarray}
&& \hspace{1.5cm}
e\frac{M_{W}^{2}}{2v^{2}}\{(k+k')_{\beta}(\theta^{\beta\gamma}g^{\sigma\rho}+\theta^{\beta\sigma}g^{\gamma\rho})+2(k_{3}^{\gamma}\theta^{\sigma\rho}\
\nonumber
\\
&& \hspace{1.5cm}
+k_{3}^{\sigma}\theta^{\gamma\rho})+2k_{3\beta}(\theta^{\rho\beta}g^{\sigma\gamma}+\theta^{\beta\gamma}g^{\sigma\rho}+\theta^{\beta\sigma}g^{\gamma\rho})\},
\end{eqnarray}
\\
\\
\\
Equation (\ref{hzzz}):
\item \hspace{-1cm}
\\
\\
\begin{picture}(55,45) (30,-30)
\SetWidth{0.5}
\DashLine(27,-10)(50,-10){3}
\Photon(70,10)(50,-10){2}{7}
\Photon(80,-10)(50,-10){2}{7}
\Photon(70,-30)(50,-10){2}{7}
\Vertex(50,-10){1.5}
\Text(12,-5)[lb]{$H(k)$}
\Text(75,8)[lb]{$Z_{\gamma}(k_1)$}
\Text(85,-15)[lb]{$Z_{\sigma}(k_2)$}
\Text(75,-44)[lb]{$Z_{\rho}(k_3)$}
\end{picture}
\vspace{1cm}
\begin{eqnarray}
&&
\hspace{1.5cm}
\frac{M_{Z}^{3}}{v^{2}}\{k_{\beta}(\theta^{\rho\beta}g^{\gamma\sigma}+\theta^{\sigma\beta}g^{\rho\gamma}+\theta^{\gamma\beta}g^{\rho\sigma})
+2k_{1\beta}(\theta^{\rho\beta}g^{\gamma\sigma}+\theta^{\sigma\beta}g^{\gamma\rho})
\nonumber
\\
&&
\hspace{1.5cm}
+2k_{2\beta}(\theta^{\rho\beta}g^{\gamma\sigma}+\theta^{\gamma\beta}g^{\sigma\rho})
+2k_{3\beta}(\theta^{\gamma\beta}g^{\sigma\rho}+\theta^{\sigma\beta}g^{\rho\gamma})
\nonumber
\\
&& \hspace{1.5cm}
+(k_{1}-k_{3})^{\sigma}\theta^{\gamma\rho}+(k_{3}-k_{2})^{\gamma}\theta^{\rho\sigma}+(k_{2}-k_{1})^{\rho}\theta^{\sigma\gamma}\},
\end{eqnarray}
\\
\\
\\
\\
Equation (\ref{hhzzz}):
\\
\item \hspace{-1cm}
\\
\\
\\
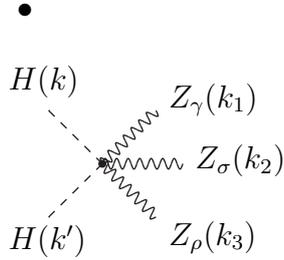
\begin{picture}(55,45) (30,-30)
\SetWidth{0.5}
\DashLine(30,10)(50,-10){3}
\DashLine(30,-30)(50,-10){3}
\Photon(70,10)(50,-10){2}{7}
\Photon(80,-10)(50,-10){2}{7}
\Photon(70,-30)(50,-10){2}{7}
\Vertex(50,-10){1.5}
\Text(15,15)[lb]{$H(k)$}
\Text(15,-45)[lb]{$H(k')$}
\Text(75,8)[lb]{$Z_{\gamma}(k_1)$}
\Text(85,-15)[lb]{$Z_{\sigma}(k_2)$}
\Text(75,-44)[lb]{$Z_{\rho}(k_3)$}
\end{picture}
\\
\vspace{.5cm}
\begin{eqnarray}
&&
\hspace{1.5cm}
\frac{M_{Z}^{3}}{v^{3}}\{(k+k')_{\beta}(\theta^{\rho\beta}g^{\gamma\sigma}+\theta^{\sigma\beta}g^{\rho\gamma}+\theta^{\gamma\beta}g^{\rho\sigma})+
2k_{1\beta}(\theta^{\rho\beta}g^{\gamma\sigma}+\theta^{\sigma\beta}g^{\gamma\rho})
\nonumber
\\
&&
\hspace{1.5cm}
+2k_{2\beta}(\theta^{\rho\beta}g^{\gamma\sigma}+\theta^{\gamma\beta}g^{\rho\sigma})+2k_{3\beta}(\theta^{\gamma\beta}g^{\rho\sigma}+\theta^{\sigma\beta}g^{\rho\gamma})
\nonumber
\\
&& \hspace{1.5cm}
+(k_{1}-k_{3})^{\sigma}\theta^{\gamma\rho}+(k_{3}-k_{2})^{\gamma}\theta^{\rho\sigma}+(k_{2}-k_{1})^{\rho}\theta^{\sigma\gamma}\},
\end{eqnarray}
\\
\\
\\
\\
\\
\\
\\
\\
Equation (\ref{hwwz}):
\item \hspace{-1cm}
\\
\\
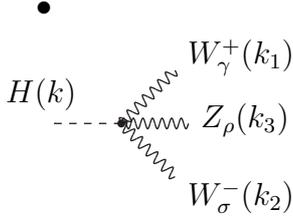
\begin{picture}(55,45) (30,-30)
\SetWidth{0.5}
\DashLine(25,-10)(50,-10){3}
\Photon(70,10)(50,-10){2}{7}
\Photon(75,-10)(50,-10){2}{7}
\Photon(70,-30)(50,-10){2}{7}
\Vertex(50,-10){1.5}
\Text(7,-5)[lb]{$H(k)$}
\Text(75,8)[lb]{$W^+_{\gamma}(k_1)$}
\Text(80,-15)[lb]{$Z_{\rho}(k_3)$}
\Text(75,-44)[lb]{$W^-_{\sigma}(k_2)$}
\end{picture}
\vspace{1cm}
\begin{eqnarray}
&&
\hspace{1.5cm}
\frac{M_{Z}M_{W}^{2}}{v^{2}}\{k_{\beta}g^{\gamma\sigma}\theta^{\rho\beta}+sin^{2}\theta_{W}(g^{\sigma\rho}\theta^{\gamma\beta}+g^{\gamma\rho}\theta^{\sigma\beta})k_{\beta}-k_{2\beta}
\nonumber
\\
&&
\hspace{1.5cm}
\times(g^{\gamma\rho}\theta^{\sigma\beta}-g^{\sigma\gamma}\theta^{\rho\beta}-g^{\rho\sigma}\theta^{\gamma\beta})+k_{1\beta}(g^{\sigma\gamma}\theta^{\rho\beta}+g^{\rho\gamma}\theta^{\sigma\beta}
\nonumber
\\
&&
\hspace{1.5cm}
-g^{\rho\sigma}\theta^{\gamma\beta})-\theta^{\gamma\sigma}(k_{2}-k_{1})^{\rho}+k_{1}^{\sigma}\theta^{\gamma\rho}+k_{2}^{\gamma}\theta^{\sigma\rho}+cos2\theta_{W}
\nonumber
\\
&& \hspace{1.5cm}
\times[k_{3}^{\gamma}\theta^{\sigma\rho}+k_{3}^{\sigma}\theta^{\gamma\rho}+k_{3\beta}(g^{\gamma\sigma}\theta^{\rho\beta}-g^{\sigma\rho}\theta^{\gamma\beta}-g^{\rho\gamma}\theta^{\sigma\beta})]\},
\end{eqnarray}
\\
\\
Equation (\ref{hhwwz}):
\item \hspace{-1cm}
\\
\\
\\
\\
\begin{picture}(55,45) (30,-30)
\SetWidth{0.5}
\DashLine(30,10)(50,-10){3}
\DashLine(30,-30)(50,-10){3}
\Photon(70,10)(50,-10){2}{7}
\Photon(75,-10)(50,-10){2}{7}
\Photon(70,-30)(50,-10){2}{7}
\Vertex(50,-10){1.5}
\Text(15,15)[lb]{$H(k)$}
\Text(15,-45)[lb]{$H(k')$}
\Text(75,8)[lb]{$W^+_{\gamma}(k_1)$}
\Text(80,-15)[lb]{$Z_{\rho}(k_3)$}
\Text(75,-44)[lb]{$W^-_{\sigma}(k_2)$}
\end{picture}

\vspace{.5cm}
\begin{eqnarray}
&&
\hspace{1.5cm}
\frac{M_{Z}M_{W}^{2}}{v^{3}}\{(k+k')_{\beta}(g^{\gamma\sigma}\theta^{\rho\beta}+sin^{2}\theta_{W}(g^{\sigma\rho}\theta^{\gamma\beta}+g^{\gamma\rho}\theta^{\sigma\beta}))-k_{2\beta}
\nonumber
\\
&&
\hspace{1.5cm}
\times(g^{\gamma\rho}\theta^{\sigma\beta}-g^{\sigma\gamma}\theta^{\rho\beta}-g^{\rho\sigma}\theta^{\gamma\beta})+k_{1\beta}(g^{\gamma\sigma}\theta^{\rho\beta}+g^{\rho\gamma}\theta^{\sigma\beta}-g^{\rho\sigma}\theta^{\gamma\beta})
\nonumber
\\
&&
\hspace{1.5cm}
-\theta^{\gamma\sigma}(k_{2}-k_{1})^{\rho}+k_{2}^{\gamma}\theta^{\sigma\rho}+k_{1}^{\sigma} \theta^{\gamma\rho}
+cos2\theta_{W}[k_{3}^{\gamma}\theta^{\sigma\rho}+k_{3}^{\sigma}\theta^{\gamma\rho}
\nonumber
\\
&& \hspace{1.5cm}
+k_{3\beta}(g^{\gamma\sigma}\theta^{\rho\beta}-g^{\sigma\rho}\theta^{\gamma\beta}
-g^{\rho\gamma}\theta^{\sigma\beta})]\},
\end{eqnarray}
\\
\\
Equation (\ref{hwwzg}):
\\
\\
\item \hspace{-1cm}
\\
\\
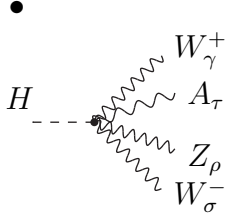
\begin{picture}(55,45) (30,-30)
\SetWidth{0.5}
\DashLine(27,-10)(50,-10){3}
\Photon(75,15)(50,-10){2}{7}
\Photon(80,1)(50,-10){2}{4}
\Photon(80,-20)(50,-10){2}{7}
\Photon(75,-35)(50,-10){2}{7}
\Vertex(50,-10){1.5}
\Text(17,-5)[lb]{$H$}
\Text(80,10)[lb]{$W^+_{\gamma}$}
\Text(85,-5)[lb]{$A_{\tau}$}
\Text(85,-30)[lb]{$Z_{\rho}$}
\Text(80,-44)[lb]{$W^-_{\sigma}$}
\end{picture}
\vspace{1cm}
\begin{eqnarray}
\frac{eM_{Z}M_{W}^{2}}{v^{2}}\{2g^{\sigma\rho}\theta^{\tau\gamma}+2g^{\gamma\rho}\theta^{\sigma\tau}+2g^{\rho\tau}\theta^{\gamma\sigma}+
g^{\tau\sigma}\theta^{\gamma\rho}+g^{\gamma\tau}\theta^{\rho\sigma}\},
\end{eqnarray}
\\
\\
Equation (\ref{hhwwzg}):
%
\item \hspace{-1cm}
\\
\\
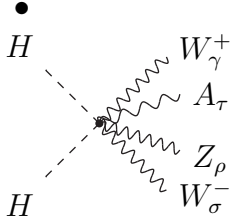
\begin{picture}(55,45) (30,-30)
\SetWidth{0.5}
\DashLine(30,10)(50,-10){3}
\DashLine(30,-30)(50,-10){3}
\Photon(75,15)(50,-10){2}{7}
\Photon(80,1)(50,-10){2}{4}
\Photon(80,-20)(50,-10){2}{7}
\Photon(75,-35)(50,-10){2}{7}
\Vertex(50,-10){1.5}
\Text(15,15)[lb]{$H$}
\Text(15,-45)[lb]{$H$}
\Text(80,10)[lb]{$W^+_{\gamma}$}
\Text(85,-5)[lb]{$A_{\tau}$}
\Text(85,-30)[lb]{$Z_{\rho}$}
\Text(80,-44)[lb]{$W^-_{\sigma}$}
\end{picture}

\begin{eqnarray}
\frac{eM_{Z}M_{W}^{2}}{v^{3}}\{2g^{\sigma\rho}\theta^{\tau\gamma}+2g^{\gamma\rho}\theta^{\sigma\tau}+
2g^{\rho\tau}\theta^{\gamma\sigma}+g^{\tau\sigma}\theta^{\gamma\rho}+g^{\gamma\tau}\theta^{\rho\sigma}\},
\end{eqnarray}
\\
\\
Equation (\ref{hwwzz}):
\item \hspace{-1cm}
\\
\\
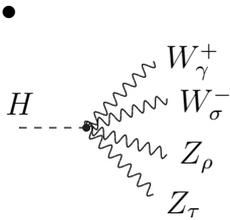
\begin{picture}(55,45) (30,-30)
\SetWidth{0.5}
\DashLine(25,-10)(50,-10){3}
\Photon(75,15)(50,-10){2}{7}
\Photon(80,1)(50,-10){2}{7}
\Photon(80,-20)(50,-10){2}{7}
\Photon(75,-35)(50,-10){2}{7}
\Vertex(50,-10){1.5}
\Text(20,-5)[lb]{$H$}
\Text(80,8)[lb]{$W^+_{\gamma}$}
\Text(85,-7)[lb]{$W^-_{\sigma}$}
\Text(85,-27)[lb]{$Z_{\rho}$}
\Text(80,-44)[lb]{$Z_{\tau}$}
\end{picture}
\vspace{1cm}
\begin{eqnarray}
&&
\hspace{1.5cm}
2\frac{M_{W}^{2}M_{Z}^{2}}{v^{3}}\{(1+\cos^{2}\theta_{W})(\theta^{\tau\gamma}g^{\sigma\rho}+\theta^{\sigma\tau}g^{\gamma\rho}
\nonumber
\\
&& \hspace{1.5cm} +
\theta^{\rho\gamma}g^{\sigma\tau}+\theta^{\sigma\rho}g^{\gamma\tau})+(1+4\cos^{2}\theta_{W})\theta^{\gamma\sigma}g^{\rho\tau}\},
\end{eqnarray}
\\
Equation (\ref{hhwwzz}):
\item \hspace{-1cm}
\\
\\
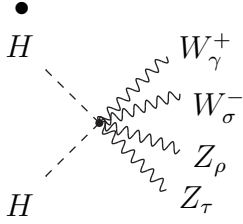
\begin{picture}(55,45) (30,-30)
\SetWidth{0.5}
\DashLine(30,10)(50,-10){3}
\DashLine(30,-30)(50,-10){3}
\Photon(75,15)(50,-10){2}{7}
\Photon(80,1)(50,-10){2}{7}
\Photon(80,-20)(50,-10){2}{7}
\Photon(75,-35)(50,-10){2}{7}
\Vertex(50,-10){1.5}
\Text(15,15)[lb]{$H$}
\Text(15,-45)[lb]{$H$}
\Text(80,10)[lb]{$W^+_{\gamma}$}
\Text(85,-10)[lb]{$W^-_{\sigma}$}
\Text(85,-30)[lb]{$Z_{\rho}$}
\Text(80,-44)[lb]{$Z_{\tau}$}
\end{picture}
\vspace{.5cm}
\begin{eqnarray}
&&
\hspace{1.5cm}
2\frac{M_{W}^{2}M_{Z}^{2}}{v^{4}}\{(1+\cos^{2}\theta_{W})(\theta^{\tau\gamma}g^{\sigma\rho}+\theta^{\sigma\tau}g^{\gamma\rho}
\nonumber
\\
&& \hspace{1.5cm} +
\theta^{\rho\gamma}g^{\sigma\tau}+\theta^{\sigma\rho}g^{\gamma\tau})+(1+4\cos^{2}\theta_{W})\theta^{\gamma\sigma}g^{\rho\tau}\},
\end{eqnarray}
\\
\\
Equation (\ref{hwwww}):
\item \hspace{-1cm}
\\
\\
\begin{picture}(55,45) (30,-30)
\SetWidth{0.5}
\DashLine(27,-10)(50,-10){3}
\Photon(75,15)(50,-10){2}{7}
\Photon(80,1)(50,-10){2}{7}
\Photon(80,-20)(50,-10){2}{7}
\Photon(75,-35)(50,-10){2}{7}
\Vertex(50,-10){1.5}
\Text(17,-5)[lb]{$H$}
\Text(80,10)[lb]{$W^+_{\gamma}$}
\Text(85,-10)[lb]{$W^+_{\tau}$}
\Text(85,-30)[lb]{$W^-_{\rho}$}
\Text(80,-44)[lb]{$W^-_{\sigma}$}
\end{picture}
\vspace{1cm}
\begin{eqnarray}
\frac{2M_{W}^{4}}{v^{3}}\{g^{\tau\sigma}\theta^{\gamma\rho}+g^{\gamma\sigma}\theta^{\tau\rho}+g^{\tau\rho}\theta^{\gamma\sigma}+
g^{\gamma\rho}\theta^{\tau\sigma}\},
\end{eqnarray}
\\
\\
Equation (\ref{hhwwww}):
\item \hspace{-1cm}
\\
\\
\begin{picture}(55,45) (30,-30)
\SetWidth{0.5}
\DashLine(30,10)(50,-10){3}
\DashLine(30,-30)(50,-10){3}
\Photon(75,15)(50,-10){2}{7}
\Photon(80,1)(50,-10){2}{7}
\Photon(80,-20)(50,-10){2}{7}
\Photon(75,-35)(50,-10){2}{7}
\Vertex(50,-10){1.5}
\Text(20,15)[lb]{$H$}
\Text(20,-45)[lb]{$H$}
\Text(80,10)[lb]{$W^+_{\gamma}$}
\Text(85,-10)[lb]{$W^+_{\tau}$}
\Text(85,-30)[lb]{$W^-_{\rho}$}
\Text(80,-44)[lb]{$W^-_{\sigma}$}
\end{picture}

\begin{eqnarray}
\frac{2M_{W}^{4}}{v^{4}}\{g^{\tau\sigma}\theta^{\gamma\rho}+g^{\gamma\sigma}\theta^{\tau\rho}+
g^{\tau\rho}\theta^{\gamma\sigma}+g^{\gamma\rho}\theta^{\tau\sigma}\}.\label{hhwwww1}
\end{eqnarray}
\\
\\


\section{Yukawa part of the NCSM Action}
\label{Yukawa}
In this section, we explore the Yukawa part of the NCSM-action up to
the first order of $\theta$ to derive all corresponding Feynman
rules. The Yukawa part of the action in terms of the physical fields
is\cite{feyn}:

\begin{eqnarray}
S_{\mbox{\tiny $\psi$, Yukawa}} & = &
\int d^4 x \sum_{i,j=1}^3
\left [
\bar{\psi}_{\mbox{\tiny down}}^{(i)}
\left(
N_{dd}^{V(ij)}
+ \gamma_5 \,
N_{dd}^{A(ij)}
\right)
\psi_{\mbox{\tiny down}}^{(j)}
\right.
\nonumber \\[0.1cm] & & \left.
+ \bar{\psi}_{\mbox{\tiny up}}^{(i)}
\left(
N_{uu}^{V(ij)}
+ \gamma_5 \,
N_{uu}^{A(ij)}
\right)
\psi_{\mbox{\tiny up}}^{(j)}
\right.
\nonumber \\[0.1cm] & & \left.
+ \bar{\psi}_{\mbox{\tiny up}}^{(i)}
\left(
C_{ud}^{V(ij)}
+ \gamma_5 \,
C_{ud}^{A(ij)}
\right)
\psi_{\mbox{\tiny down}}^{(j)}
\right.
\nonumber \\[0.1cm] & & \left.
+ \bar{\psi}_{\mbox{\tiny down}}^{(i)}
\left(
C_{du}^{V(ij)}
+ \gamma_5 \,
C_{du}^{A(ij)}
\right)
\psi_{\mbox{\tiny up}}^{(j)}
\right]
\, .
\label{eq:SYukawa-fiz}
\end{eqnarray}
The neutral currents are:
\begin{eqnarray}
N_{dd}^{V(ij)}
&=&
-M_{\mbox{\tiny down}}^{(ij)} \left(1 + \frac{h}{v}\right)
+
N_{dd}^{V,\theta (ij)}
+
{\cal O}(\theta^2)
\, ,
\nonumber \\
N_{dd}^{A(ij)}
&=&
N_{dd}^{A,\theta (ij)}
+
{\cal O}(\theta^2)
\, ,
\nonumber
\\
N_{uu}^{V(ij)}
&=&
-M_{\mbox{\tiny up}}^{(ij)} \left(1 + \frac{h}{v}\right)
+
N_{uu}^{V,\theta (ij)}
+
{\cal O}(\theta^2)
\, ,
\nonumber \\
N_{uu}^{A(ij)}
&=&
N_{uu}^{A,\theta (ij)}
+
{\cal O}(\theta^2)
\, ,
\label{eq:dV}
\end{eqnarray}
where
\begin{eqnarray}
\lefteqn{N_{dd}^{V,\theta (ij)}}
\nonumber \\
&=&
- \frac{1}{2} \theta^{\mu \nu}
M_{\mbox{\tiny down}}^{(ij)}
\left\{
i \frac{(\partial_{\mu}h)}{v} \stackrel{\rightarrow}{\partial}_{\nu}
\right.
\nonumber \\ & &
\left.
-
\left[
e Q_{\psi_{\mbox{\tiny down}}} A_{\mu}
+
\frac{g}{2 \cos \theta_W}
(T_{3,\psi_{\mbox{\tiny down},L}}-
2 Q_{\psi_{\mbox{\tiny down}}} \sin^2 \theta_W) Z_{\mu}
\right]
 \frac{(\partial_{\nu}h)}{v}
\right.
\nonumber \\ & &
\left.
+
\left[
e Q_{\psi_{\mbox{\tiny down}}} (\partial_\nu A_{\mu})
+
\frac{g}{2 \cos \theta_W}
(T_{3,\psi_{\mbox{\tiny down},L}}-
2 Q_{\psi_{\mbox{\tiny down}}} \sin^2 \theta_W)
(\partial_\nu Z_{\mu})
\right.
\right.
\nonumber \\ & &
\left.
\left.
\quad -i \, \frac{g^2}{2} W_{\mu}^+ W_{\nu}^-
\right.
\Big]
\left( 1 + \frac{h}{v}\right)
\right\}
\, ,
\label{eq:NdV}
\end{eqnarray}
\begin{eqnarray}
N_{dd}^{A,\theta (ij)}
&=&
\frac{g}{4 \cos \theta_W}
\,
T_{3,\psi_{\mbox{\tiny down},L}}
\,
 \theta^{\mu \nu}
M_{\mbox{\tiny down}}^{(ij)}
\left( 1 + \frac{h}{v}\right)
Z_{\mu}
\nonumber \\ & & \times
\left[
\left(
\stackrel{\leftarrow}{\partial}_{\nu}
- \stackrel{\rightarrow}{\partial}_{\nu}
\right)
+2 i
e
Q_{\psi_{\mbox{\tiny down}}}
A_{\nu}
\right]\, , \qquad
\label{eq:NdA}
\end{eqnarray}
and
\begin{equation}
\left.
\begin{array}{l}
 N_{uu}^{V,\theta (ij)}\\[0.2cm]
N_{uu}^{A,\theta (ij)}
\end{array}
\right\}
 =
\left\{
\begin{array}{l}
N_{dd}^{V,\theta (ij)}\\[0.2cm]
N_{dd}^{A,\theta (ij)}
\end{array}
\right.
(
W^+  \leftrightarrow  W^-,
{\mbox{down}}
\rightarrow
\mbox{up}
 ).
\label{eq:NuVA}
\end{equation}
The charged currents are given by
\begin{eqnarray}
C_{ud}^{V(ij)}
&=&
C_{ud}^{V,\theta (ij)}
+
{\cal O}(\theta^2)
\, ,
\nonumber \\[0.1cm]
C_{ud}^{A(ij)}
&=&
C_{ud}^{A,\theta (ij)}
+
{\cal O}(\theta^2)
\, ,
\end{eqnarray}
where
\begin{eqnarray}
\lefteqn{C_{ud}^{V,\theta (ij)}}
\nonumber \\
&=&
- \frac{g}{4 \sqrt{2}}
\theta^{\mu \nu}
\left( 1 + \frac{h}{v}\right)
\left\{
\left[
\left(
(V_{f} M_{\mbox{\tiny down}})^{(ij)}+
(M_{\mbox{\tiny up}} V_{f})^{(ij)}
\right)
(\partial_{\nu} W^+_{\mu})
\right.
\right.
\nonumber \\[0.1cm] & &
\left.
\left.
+
\left(
(V_{f} M_{\mbox{\tiny down}})^{(ij)}
\stackrel{\rightarrow}{\partial}_{\nu}
+
(M_{\mbox{\tiny up}} V_{f})^{(ij)}
\stackrel{\leftarrow}{\partial}_{\nu}
\right)
W^+_{\mu}
\right]
\right.
\nonumber \\[0.1cm] & &
\left.
+ i e
\left(
(V_{f} M_{\mbox{\tiny down}})^{(ij)}
Q_{\psi_{\mbox{\tiny up}}}
-
(M_{\mbox{\tiny up}} V_{f})^{(ij)}
Q_{\psi_{\mbox{\tiny down}}}
\right)
A_{\mu} W_{\nu}^+
\right.
\nonumber \\[0.1cm] & &
\left.
+ i \frac{g}{\cos \theta_W}
\left[
(V_{f} M_{\mbox{\tiny down}})^{(ij)}
\left(
2 T_{3,\psi_{\mbox{\tiny up},L}}
-
Q_{\psi_{\mbox{\tiny up}}}
\sin^2 \theta_W
\right)
\right.
\right.
\nonumber \\[0.1cm] & &
\left.
\left.
- (M_{\mbox{\tiny up}} V_{f})^{(ij)}
\left(
2 T_{3,\psi_{\mbox{\tiny down},L}}
-
Q_{\psi_{\mbox{\tiny down}}}
\sin^2 \theta_W
\right)
\right]
Z_{\mu} W_{\nu}^+
\right\}
\, ,
\end{eqnarray}
and
\begin{equation}
C_{ud}^{A,\theta (ij)} =
C_{ud}^{V,\theta (ij)}(M_{\mbox{\tiny up}} \rightarrow - M_{\mbox{\tiny up}})
\, ,
\end{equation}
while
\begin{eqnarray}
C_{du}^{V (ij)} & = &
\left( C_{ud}^{V (ij)}
(\stackrel{\rightarrow}{\partial}
\leftrightarrow
\stackrel{\leftarrow}{\partial}
)\right) ^{\dagger}
\, ,
\nonumber \\
C_{du}^{A (ij)} & = & - \left( C_{ud}^{A (ij)}
(\stackrel{\rightarrow}{\partial} \leftrightarrow
\stackrel{\leftarrow}{\partial} )\right) ^{\dagger}\, ,
\label{eq:ChVA}
\end{eqnarray}
where $\stackrel{\rightarrow}{\partial}_{\rho}$
($\stackrel{\leftarrow}{\partial}_{\rho}$ ) denotes the partial
derivative which acts only on the fermion fields on its right (left)
side:
\begin{equation}
{\partial}_{\rho} \psi \equiv
\stackrel{\rightarrow}{\partial}_{\rho} \psi
 \qquad
{\partial}_{\rho} \overline{\psi} \equiv
\overline{\psi} \stackrel{\leftarrow}{\partial}_{\rho}.
\end{equation}
Now, the Yukawa-action (\ref{eq:SYukawa-fiz}) can be simplified to
find all lagrangian densities
containing the Higgs couplings with fermions as follows:
\\
\\
\\
\item

\begin{itemize}
\item Higgs-Fermion-Antifermion:
\begin{equation}
\hspace{1.5cm}
 \mathcal{L}_{h\bar{f}f}=\frac {- m_{f}}{v}\bar{f}(h+\frac{i}{2}\theta^{\mu\nu}(\partial_{\mu}h)\stackrel{\rightarrow}\partial_{\nu})
 f,
 \label{hff}
\end{equation}
\item Higgs-Photon-Fermion-Antifermion:
\begin{equation}
\hspace{1.5cm}
 \mathcal{L}_{h\bar{f}fA}= \frac {eQ_{f}m_{f}}{2v}\theta^{\mu\nu}[A_{\mu}(\partial_{\nu}h)-h(\partial_{\nu}A_{\mu})]
 \bar{f}f,
 \label{hgff}
\end{equation}
\item Higgs-Z boson-Fermion-Antifermion:
\begin{eqnarray}
\hspace{1.5cm}
 \mathcal{L}_{h\bar{f}fZ} &=&
  \frac {g m_{f}}{4v cos\theta_{W}}\theta^{\mu\nu}[c_{V,f}Z_{\mu}(2(\partial_{\nu}h)\bar{f}f +
 \nonumber \\ &&
 +h(\partial_{\nu}\bar{f})f+h(\partial_{\nu}f)\bar{f})+\gamma^5c_{A,f}h((\partial_{\nu}\bar{f})f
 \nonumber \\ &&
 -(\partial_{\nu}f)\bar{f})Z_{\mu}],
 \label{hzff}
\end{eqnarray}
\item Higgs-Z boson-Photon-Fermion-Antifermion:
\begin{eqnarray}
\hspace{1.5cm}
 \mathcal{L}_{h\bar{f}fZA} =\frac {ie\gamma^{5}Q_{f}g m_{f}}{2 v cos\theta_{W}}\theta^{\mu\nu}c_{A,f}h
 Z_{\mu}A_{\nu}\bar{f}f,
  \label{hzgff}
\end{eqnarray}
\item Higgs-$W^{+}$ boson-$W^{-}$ boson-Fermion-Antifermion:
\begin{eqnarray}
\hspace{1.5cm}
 \mathcal{L}_{h\bar{f}fW^{+}W^{-}} =\frac {im_{f}g^{2}}{4v}\theta^{\mu\nu}h
 f\bar{f}W^{+}_{\mu}W^{-}_{\nu},
  \label{hwwff}
\end{eqnarray}
\item Higgs-$W^{+}$ boson-d-$\bar{u}$:
\begin{eqnarray}
\hspace{1.5cm}
 \mathcal{L}_{hW^{+}d \bar{u}} &=& -\frac{g}{4\sqrt{2}v}\theta^{\mu\nu}V_{f}^{ij}h\bar{u}^{i}\{(m_{d^{j}}+m_{u^{i}})
  \nonumber \\&&
 \times(\partial_{\nu} W^{+}_{\mu})+(m_{d^{j}}\stackrel{\rightarrow}{\partial_{\nu}}
 +m_{u^{i}}\stackrel{\leftarrow}{\partial_{\nu}})
 \nonumber\\&&
  \times W^{+}_{\mu}+\gamma^{5}[(m_{d^{j}}-m_{u^{i}}) (\partial_{\nu}W^{+}_{\mu})
   \nonumber \\&&
   +(m_{d^{j}}\stackrel{\rightarrow}{\partial_{\nu}}
 -m_{u^{i}}\stackrel{\leftarrow}{\partial_{\nu}})W^{+}_{\mu}]\}d^{j},
  \label{hwdu}
\end{eqnarray}
\item Higgs-$W^{-}$ boson-u-$\bar{d}$:
\begin{eqnarray}
\hspace{1.5cm}
 \mathcal{L}_{hW^{-}u \bar{d}} &=& -\frac{g}{4\sqrt{2}v}\theta^{\mu\nu}V_{f}^{\ast ij}\bar{d}^{j}\{(m_{d^{j}}+m_{u^{i}})
  \nonumber \\&&
 \times(\partial_{\nu} W^{-}_{\mu})+(m_{d^{j}}\stackrel{\leftarrow}{\partial_{\nu}}
 +m_{u^{i}}\stackrel{\rightarrow}{\partial_{\nu}})
 \nonumber\\&&
 \times W^{-}_{\mu} -\gamma^{5}[(m_{d^{j}}-m_{u^{i}}) (\partial_{\nu}W^{-}_{\mu})
   \nonumber \\&&
   +(m_{d^{j}}\stackrel{\leftarrow}{\partial_{\nu}}
 -m_{u^{i}}\stackrel{\rightarrow}{\partial_{\nu}})W^{-}_{\mu}]\}hu^{i},
  \label{hwud}
\end{eqnarray}
\item Higgs-$W^{+}$ boson-photon-d-$\bar{u}$:
\begin{eqnarray}
\hspace{1.5cm}
 \mathcal{L}_{hW^{+}Ad\bar{u}} &=& -\frac{ie g}{4\sqrt{2}v}\theta^{\mu\nu}V_{f}^{ij}\bar{u}^{i}\{(m_{d^{j}}Q_{u^{i}}-m_{u^{i}}Q_{d^{j}})
  \nonumber \\&&
   +\gamma^{5}(m_{d^{j}}Q_{u^{i}}+m_{u^{i}}Q_{d^{j}})\}
   A_{\mu}W^{+}_{\nu}hd^{j},  \label{hwgdu}
  \end{eqnarray}
\item Higgs-$W^{-}$ boson-photon-u-$\bar{d}$:
\begin{eqnarray}
\hspace{1.5cm}
 \mathcal{L}_{hW^{-}Au\bar{d}} &=& \frac{ie g}{4\sqrt{2}v}\theta^{\mu\nu}V_{f}^{\ast ij}\bar{d}^{j}\{(m_{d^{j}}Q_{u^{i}}-m_{u^{i}}Q_{d^{j}})
  \nonumber \\&&
   -\gamma^{5}(m_{d^{j}}Q_{u^{i}}+m_{u^{i}}Q_{d^{j}})\}
   A_{\mu}W^{-}_{\nu}hu^{i},  \label{hwgud}
  \end{eqnarray}
\item Higgs-$W^{+}$ boson-Z boson-d-$\bar{u}$:
  \begin{eqnarray}
\hspace{1.5cm}
 \mathcal{L}_{hW^{+}Zd\bar{u}} &=& -\frac{ig^2 }{8\sqrt{2}vcos\theta_{W}}\theta^{\mu\nu}V_{f}^{ ij}\{m_{d^{j}}(3c_{A,f}+c_{V,f})(1+\gamma^{5})
 \nonumber \\&&
 -m_{f_{i}}(3c_{A,f}+c_{V,f})(1-\gamma^{5})
   \}\bar{u}^{i}Z_{\mu}W^{+}_{\nu}hd^{j},  \label{hwzdu}
  \end{eqnarray}
\item Higgs-$W^{-}$ boson-Z boson-u-$\bar{d}$:
  \begin{eqnarray}
\hspace{1.5cm}
 \mathcal{L}_{hW^{-}Zu\bar{d}} &=& \frac{ig^2}{8\sqrt{2}vcos\theta_{W}}\theta^{\mu\nu}V_{f}^{\ast ij}
  \{m_{d^{j}}(3c_{A,f}+c_{V,f})(1-\gamma^{5})
 \nonumber \\&&
 -m_{u^{i}}(3c_{A,f}+c_{V,f})(1+\gamma^{5})
  \}\bar{d}^{j}Z_{\mu}W^{-}_{\nu}hu^{i},  \label{hwzud}
  \end{eqnarray}
\end{itemize}
where $u$ and $d$ indicate $\{u,c,t\}$ and $\{d,s,b\}$,
respectively, and
\begin{eqnarray}
c_{V,f} & = & T_{3,f_{L}}\, -\,  2\,  Q_{f}\,  \sin^2 \theta_W\,,
\nonumber\\
c_{A,f} & = & T_{3,f_L} \,. \label{cva}
\end{eqnarray}
\subsection{\label{Y}{Feynman Rules for Yukawa Terms}}
In this subsection, we list the Feynman rules for the Yukawa terms
of the NCSM action up to the first order of $\theta$:
\\
\\
Equation (\ref{hff}):
\item \hspace{-1cm}
\\
\\
\begin{picture}(55,45) (30,-30)
\SetWidth{0.5}
\ArrowLine(70,-25)(50,-10)
\ArrowLine(50,-10)(70,6)
\DashLine(50,-10)(20,-10){3}
\Vertex(50,-10){1.5}
\Text(60,-40)[lb]{$f$}
\Text(60,10)[lb]{$f$}
\Text(15,-8)[lb]{$H(k)$}
\end{picture}
\\
\vspace{-1.5cm}
\begin{equation}
\hspace{1.5cm}
-\frac{im_{f}}{v}(1-\frac{i}{2}\theta_{\mu\nu}k^{\mu}k_{in}^{\nu}),
\label{hff1}
\end{equation}
\vspace{1cm}
\\
Equation (\ref{hgff}):
\item \hspace{-1cm}
\\
\\
\begin{picture}(55,45) (30,-30)
\SetWidth{0.5}
\ArrowLine(70,-30)(50,-10)
\ArrowLine(50,-10)(70,11)
\DashLine(50,-10)(20,-10){3}
\Photon(80,-10)(50,-10){2}{4}
\Vertex(50,-10){1.5}
\Text(60,-40)[lb]{$f$}
\Text(83,-10)[lb]{$A_{\mu}(k')$}
\Text(60,10)[lb]{$f$}
\Text(15,-8)[lb]{$H(k)$}
\end{picture}
%
%
\begin{equation}
\hspace{1.5cm}
\frac{eQ_{f}}{2v}m_{f}\theta^{\mu\nu}(k-k')_{\nu},
\end{equation}
\\
\\
 Equation (\ref{hzff}):
\item \hspace{-1cm}
\\
\\
\begin{picture}(55,45) (30,-30)
\SetWidth{0.5}
\ArrowLine(70,-30)(50,-10)
\ArrowLine(50,-10)(70,11)
\DashLine(50,-10)(20,-10){3}
\Photon(80,-10)(50,-10){2}{7}
\Vertex(50,-10){1.5}
\Text(60,-40)[lb]{$f$}
\Text(83,-10)[lb]{$Z_{\mu}(k')$}
\Text(60,10)[lb]{$f$}
\Text(15,-8)[lb]{$H(k)$}
\end{picture}
\\
\\
\begin{equation}
\hspace{1.5cm}
\frac{m_{f}M_{Z}}{2v^{2}}\theta^{\mu\nu}[c_{V,f}(2k+k_{in}-k_{out})_{\nu}
\, -\gamma_{5}c_{A,f}(k_{in}+k_{out})_{\nu}],
\end{equation}
\\
\vspace{1cm}
Equation (\ref{hzgff}):
\item \hspace{-1cm}
\\
\\
\begin{picture}(55,45) (30,-30)
\SetWidth{0.5}
\ArrowLine(70,-35)(50,-10)
\ArrowLine(50,-10)(70,15)
\DashLine(50,-10)(20,-10){3}
\Photon(80,4)(50,-10){2}{4}
\Photon(80,-24)(50,-10){2}{7}
\Vertex(50,-10){1.5}
\Text(65,-50)[lb]{$f$}
\Text(83,2)[lb]{$A_{\nu}$}
\Text(83,-27)[lb]{$Z_{\mu}$}
\Text(65,20)[lb]{$f$}
\Text(15,-8)[lb]{$H$}
\end{picture}

\vspace{-2.cm} 
\begin{equation}
\hspace{1.5cm}
-\frac{em_{f}M_{Z}}{v^{2}}Q_{f}c_{A,f}\gamma_{5}\theta^{\mu\nu},
\end{equation}
\\
\\
\\
\\
\\
\\
\\
\\
\\
Equation (\ref{hwwff}):
\item\hspace{1cm}
\\
\\
\hspace*{1cm}
\begin{picture}(55,45) (30,-30)
\SetWidth{0.5}
\ArrowLine(70,-35)(50,-10)
\ArrowLine(50,-10)(70,15)
\DashLine(50,-10)(20,-10){3}
\Photon(80,4)(50,-10){2}{7}
\Photon(80,-24)(50,-10){2}{7}
\Vertex(50,-10){1.5}
\Text(65,-50)[lb]{$f_{u}$}
\Text(83,2)[lb]{$W^{+}_{\mu}$}
\Text(83,-27)[lb]{$W^{-}_{\nu}$}
\Text(65,20)[lb]{$f_{u}$}
\Text(20,-8)[lb]{$H$}
\end{picture}
\hspace*{3cm}
\begin{picture}(55,45) (30,-30)
\SetWidth{0.5}
\ArrowLine(70,-35)(50,-10)
\ArrowLine(50,-10)(70,15)
\DashLine(50,-10)(20,-10){3}
\Photon(80,4)(50,-10){2}{7}
\Photon(80,-24)(50,-10){2}{7}
\Vertex(50,-10){1.5}
\Text(65,-50)[lb]{$f_{d}$}
\Text(83,2)[lb]{$W^{+}_{\mu}$}
\Text(83,-27)[lb]{$W^{-}_{\nu}$}
\Text(65,20)[lb]{$f_{d}$}
\Text(20,-8)[lb]{$H$}
\end{picture}
\vspace{2cm} 
\begin{equation}
\hspace{-1cm} \frac{M_{W}^{2}m_{f}}{v^{3}}\theta^{\mu\nu},
\hspace{3cm} -\frac{M_{W}^{2}m_{f}}{v^{3}}\theta^{\mu\nu},
\end{equation}
\\
Equation (\ref{hwdu}):
\item \hspace{-1cm}
\\
\\
\begin{picture}(55,45) (30,-30)
\SetWidth{0.5}
\ArrowLine(50,-10)(70,-30)
\ArrowLine(70,11)(50,-10)
\DashLine(50,-10)(20,-10){3}
\Photon(80,-10)(50,-10){2}{7}
\Vertex(50,-10){1.5}
\Text(60,-45)[lb]{$u^{i}$}
\Text(83,-10)[lb]{$W^{+}_{\mu}(k')$}
\Text(60,10)[lb]{$d^{j}$}
\Text(15,-8)[lb]{$H(k)$}
\end{picture}
\\
\vspace{1.cm}
\begin{equation}
\hspace{1.5cm} \vspace{1.cm}
-\frac{M_{W}}{2\sqrt{2}v^{2}}V^{ij}_{f}\theta^{\mu\nu}[m_{f_{d}^{j}}(k'+k_{in})_{\nu}(1+\gamma_{5})\,
+m_{f_{u}^{i}}(k'-k_{out})_{\nu}(1-\gamma_{5})],
\end{equation}
Equation (\ref{hwud}):
\item \hspace{-1cm}
\\
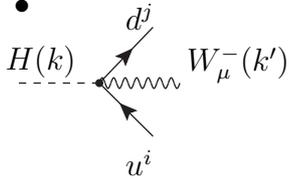
\begin{picture}(55,45) (30,-30)
\SetWidth{0.5}
\ArrowLine(70,-30)(50,-10)
\ArrowLine(50,-10)(70,11)
\DashLine(50,-10)(20,-10){3}
\Photon(80,-10)(50,-10){2}{7}
\Vertex(50,-10){1.5}
\Text(60,-45)[lb]{$u^{i}$}
\Text(83,-10)[lb]{$W^{-}_{\mu}(k')$}
\Text(60,10)[lb]{$d^{j}$}
\Text(15,-8)[lb]{$H(k)$}
\end{picture}
\\
\\
\begin{equation}
\hspace{1cm} -\frac{M_{W}}{2\sqrt{2}v^{2}}V^{\ast
ij}_{f}\theta^{\mu\nu}[m_{f_{d}^{j}}(k'-k_{out})_{\nu}(1-
\gamma_{5})+m_{f_{u}^{i}}(k'+k_{in})_{\nu}(1+\gamma_{5})],
\end{equation}
\\
\\
\vspace{.7cm}
Equation (\ref{hwgdu}):
\item \hspace{-1cm}
\\
\begin{picture}(55,45) (30,-30)
\SetWidth{0.5}
\ArrowLine(70,-35)(50,-10)
\ArrowLine(50,-10)(70,15)
\DashLine(50,-10)(20,-10){3}
\Photon(80,4)(50,-10){2}{4}
\Photon(80,-24)(50,-10){2}{7}
\Vertex(50,-10){1.5}
\Text(65,-50)[lb]{$d^{j}$}
\Text(83,2)[lb]{$A_{\mu}$}
\Text(83,-27)[lb]{$W^{+}_{\nu}$}
\Text(65,20)[lb]{$u^{i}$}
\Text(15,-8)[lb]{$H(k)$}
\end{picture}
\\
\vspace{1.cm} 
\begin{equation}
\hspace{1.5cm}
\frac{eM_{W}}{2\sqrt{2}v^{2}}V^{ij}_{f}\theta^{\mu\nu}[m_{f_{d}^{j}}Q_{f_{u}^{i}}(1+
\gamma_{5})-m_{f_{u}^{i}}Q_{f_{d}^{j}}(1-\gamma_{5})],
\end{equation}
\\
Equation (\ref{hwgud}):
\item \hspace{-1cm}
\\
\\
\begin{picture}(55,45) (30,-30)
\SetWidth{0.5}
\ArrowLine(70,-35)(50,-10)
\ArrowLine(50,-10)(70,15)
\DashLine(50,-10)(20,-10){3}
\Photon(80,4)(50,-10){2}{7}
\Photon(80,-24)(50,-10){2}{4}
\Vertex(50,-10){1.5}
\Text(65,-50)[lb]{$u^{i}$}
\Text(83,2)[lb]{$W^{-}_{\nu}$}
\Text(83,-27)[lb]{$A_{\mu}$}
\Text(65,20)[lb]{$d^{j}$}
\Text(15,-8)[lb]{$H(k)$}
\end{picture}
\vspace{1.cm}
\begin{equation}
\hspace{1.5cm} -\frac{eM_{W}}{2\sqrt{2}v^{2}}V^{\ast
ij}_{f}\theta^{\mu\nu}[m_{f_{d}^{j}}Q_{f_{u}^{i}}(1-
\gamma_{5})-m_{f_{u}^{i}}Q_{f_{d}^{j}}(1+\gamma_{5})],
\end{equation}
\\
Equation (\ref{hwzdu}):
\item \hspace{-1cm}
\\
\\
\begin{picture}(55,45) (30,-30)
\SetWidth{0.5}
\ArrowLine(70,-35)(50,-10)
\ArrowLine(50,-10)(70,15)
\DashLine(50,-10)(20,-10){3}
\Photon(80,4)(50,-10){2}{7}
\Photon(80,-24)(50,-10){2}{7}
\Vertex(50,-10){1.5}
\Text(65,-50)[lb]{$d^{j}$}
\Text(83,2)[lb]{$W^{+}_{\nu}$}
\Text(83,-27)[lb]{$Z_{\mu}$}
\Text(65,20)[lb]{$u^{i}$}
\Text(15,-8)[lb]{$H(k)$}
\end{picture}
\vspace{1.cm}
\begin{equation}
\hspace{1.5cm}
\frac{M_{Z}M_{W}}{2\sqrt{2}v^{3}}V^{ij}_{f}\theta^{\mu\nu}[m_{f_{d}^{j}}(c_{V,f}+3c_{A,f})_{u}(1+
\gamma_{5})-m_{f_{u}^{i}}(c_{V,f}+3c_{A,f})_{d}(1-\gamma_{5})],
\end{equation}
\\
\\
\\
\\
\\
Equation (\ref{hwzud}):
\item \hspace{-1cm}
\\
\\
\begin{picture}(55,45) (30,-30)
\SetWidth{0.5}
\ArrowLine(50,-10)(70,-35)
\ArrowLine(70,15)(50,-10)
\DashLine(50,-10)(20,-10){3}
\Photon(80,4)(50,-10){2}{7}
\Photon(80,-24)(50,-10){2}{7}
\Vertex(50,-10){1.5}
\Text(65,-50)[lb]{$d^{j}$}
\Text(83,2)[lb]{$W^{-}_{\nu}$}
\Text(83,-27)[lb]{$Z_{\mu}$}
\Text(65,20)[lb]{$u^{i}$}
\Text(15,-8)[lb]{$H(k)$}
\end{picture}
\vspace{1.cm}
\begin{equation}
\hspace{1.5cm} -\frac{M_{Z}M_{W}}{2\sqrt{2}v^{3}}V^{\ast
ij}_{f}\theta^{\mu\nu}[m_{f_{d}^{j}}(c_{V,f}+3c_{A,f})_{u}(1-
\gamma_{5})-m_{f_{u}^{i}}(c_{V,f}+3c_{A,f})_{d}(1+\gamma_{5})],
\label{hwzud1}
\end{equation}

\section{\label{f}{Summery}}
We examined the Higgs and Yukawa parts of the NCSM-action to find
all the Higgs couplings with gauge and fermion fields, see
(\ref{hzz})-(\ref{hhwwww}) and (\ref{hff})-(\ref{hwzud}),
respectively. We obtained the corresponding Feynman rules as is
given in (\ref{hhz1})-(\ref{hhwwww1}) and
(\ref{hff1})-(\ref{hwzud1}). One can easily see that besides the
usual standard model interactions, there are new couplings between
the Higgs and fermions and the electroweak gauge bosons such as
$Zhh$, $Zhhh$, $Zhhhh$, $hhhW^+W^-$, $h\gamma W^+W^-$, $hZ W^+W^-$,
$ZZZh$, $h\gamma ZW^+W^-$,..., from the Higgs part and $h{\bar
f}f\gamma$, $h{\bar f}fZ$, $h{\bar f}f\gamma Z$, $h{\bar f}fW^+W^-$,
$h{\bar u}d W^+$, $h{\bar u}d W^+\gamma$, $h{\bar u}d W^+Z$ and so
on from the Yukawa part. These new vertices, if exist, lead to new
production and decay channels that will be very important to
consider in LHC or the future ILC colliders.   For instance, in the
decay of Higgs
 to four leptons, $H \rightarrow llll$, which is one of the main
discovery channels for the Higgs boson \cite{CMS}, there are new
vertices in the NCSM such as $h{\bar f}f\gamma$ and $h{\bar f}fZ$.
Meanwhile, in the NC space the Higgs can be produced through new
channels such as $q q \rightarrow WW^{*} \rightarrow q q H \gamma$,
$q q \rightarrow WW^{*} \rightarrow q q H Z$, $q q \rightarrow
ZZ^{*} \rightarrow q q H Z$, and so on.

\end{itemize}

\end{document}